\documentclass[12pt]{JHEP3}

\usepackage{amsmath}
\newcommand{\be}[1]{ \begin{equation}\label{#1} }
\newcommand{\ee}{\end{equation}}
\newcommand{\bea}[1]{\begin{eqnarray}\label{#1} }
\newcommand{\eea}{\end{eqnarray}}

\def\ZZZ{{\hskip-3pt\hbox{ Z\kern-1.6mm Z}}}
\def\zzz{{\hskip-3pt\hbox{ z\kern-1mm z}}}
\usepackage{amstext,amssymb,amsfonts}

\newcommand{\ads}{{\text{AdS}_5}}

\def\one{{\hbox{ 1\kern-.8mm l}}}
\def\zero{{\hbox{ 0\kern-1.5mm 0}}}

\def\ZZ{{\mathcal Z}}
\def\YY{{\mathcal Y}}

\def\bz{{\mathbb Z}}
\def\tr{{\rm Tr}}

\def\l{\left}
\def\r{\right}

\def\D{\Delta}

\def\g{\gamma}

\def\a{\alpha}

\def\o[#1]{{\rm O}\left({#1}\right)}
\def\dotl[#1,#2]{\left\langle #1, #2 \right\rangle}
\def\dotlb[#1,#2]{[ #1, #2 ]}
\def\dotp[#1,#2]{(#1) \cdot (#2)}
\def\>{\rangle}
\def\<{\langle}

\newcommand{\ben}[1]{\begin{eqnarray}\label{#1} }
\newcommand{\een}{\end{eqnarray}}

\def\r{\right}
\def\l{\left}

\title{CFT(4) Partition Functions and the Heat Kernel on AdS(5)}

\author{Shailesh Lal\footnote{shailesh DOT lal AT icts DOT res DOT in} 
\\
International Center for Theoretical Sciences -- TIFR, \\
TIFR Centre Building, Indian Institute of Science,\\
Bangalore, India 560012.\\
}

\abstract{We explicitly reorganise the partition function of an arbitrary CFT in four spacetime dimensions into a heat kernel form for the dual string spectrum on AdS(5). On very general grounds, the heat kernel answer can be expressed in terms of a convolution of the one-particle partition function of the four-dimensional CFT. Our methods are general and would apply for arbitrary dimensions, which we comment on.}
\preprint{ICTS/2012/12}

\begin{document}

\baselineskip 3.5ex

\section{Introduction}
Heat kernel methods have of late played an important role in extracting out quantum effects in gravitational physics. Such applications include the extraction of leading quantum corrections to black hole entropy \cite{Banerjee:2010qc}--\cite{Sen:2012dw}, the asymptotic symmetries of gravitational theories \cite{Giombi:2008vd}--\cite{Gupta:2012np}, precision tests of AdS/CFT \cite{Bhattacharyya:2012ye}, as well as the many more applications extensively reviewed in \cite{Vassilevich:2003xt}. This is essentially because the heat kernel method powerfully captures the leading quantum properties of a given theory. In many cases (especially quantum gravity) while the full quantum theory is poorly understood, these leading properties are potentially tractable.

In this paper, we shall briefly describe some progress in bringing heat kernel methods to bear on another significant arena, the string theory sigma model on AdS. As is well known, this sigma model is presently quite intractable at the quantum level. In this light, one potential starting point to gain a foothold on the sigma model could be to use the AdS/CFT correspondence \cite{Maldacena:1997re,Gubser:1998bc,Witten:1998qj}. In particular, to take the spectrum of the CFT dual to the AdS string and to reproduce the planar CFT partition function in terms of quadratic fluctuations of the dual fields in AdS\footnote{For a free planar CFT defined on the compact space $S^3\otimes S^1$ this partition function is very explicitly known by counting the spectrum of gauge invariant operators \cite{Sundborg:1999ue,Aharony:2003sx}, see also \cite{Polyakov:2001af} and the related papers \cite{Bianchi:2003wx}--\cite{Newton:2008au}.}. Typically, these quadratic fluctuations would arrange themselves into determinants of the Laplacian acting over fields of varying spin. This would essentially provide us with a first-quantised description of the particles that form the string spectrum. More ambitiously, one could attempt to reconstruct the full vacuum amplitude (the torus string amplitude with no vertex operator insertions) in AdS by interpreting the heat kernel proper time in terms of the modulus of the torus worldsheet \cite{David:2009xg}\footnote{There are good reasons to expect this approach to be a fruitful one. In particular, it has been proposed that the general relation between the heat kernel proper time and the closed string moduli is closely connected to the phenomenon of gauge-string duality \cite{Gopakumar:2003ns}--\cite{Gopakumar:2005fx}, see also \cite{Mamedov:2005uw} for an interesting application of this approach. Encouragingly, this overall approach based on the heat kernel method has also previously been successful for string theory in flat space \cite{Polchinski:1985zf}. We thank Rajesh Gopakumar for helpful discussions and correspondence about these points.}.

In this paper, we shall perform the first of these two tasks. We shall organise the CFT partition function in terms of quadratic fluctuations of the particles that constitute the string spectrum on AdS. We shall do this in the specific instance of the $\ads$/CFT$_4$ duality. However, we make no assumptions about the matter content of the CFT or about supersymmetry. 

A brief overview of this paper is as follows. In Section \ref{seckernel} we will begin with a brief review of the heat kernel methods of \cite{Gopakumar:2011qs} and their subsequent applications in \cite{Gupta:2012np}. We shall then present our ansatz for the extension of these results to (massive or massless) fields of mixed symmetry in Section \ref{secansatz}. Finally, in Section \ref{secanswers}, we use the above results to interpret the given CFT partition function in terms of the heat kernel of the dual AdS(5) spectrum. The equation \eqref{longfinal} obtained there is the central result of this paper. Subtleties that occur due to the appearance of short multiplets in the CFT operator spectrum are relegated to the Appendix \ref{appshortkernel} as they do not affect the final results.
\section{The Heat Kernel for AdS(5): A Review}\label{seckernel}
The essential ingredient that we employ to compute the one-loop partition function in the bulk is its relation to the determinant of the Laplacian, which in turn may be evaluated from the (traced, coincident) heat kernel. In this section, we shall review the results of \cite{Gopakumar:2011qs} for the heat kernel of the Laplacian acting over tensor fields on AdS. In summary, for a spin-S particle moving on a spacetime manifold $\mathcal{M}$,
\begin{equation}
\ln\ZZ^{(S)}\simeq \ln\det\l(-\nabla^2_{(S)}\r)=\tr\ln\l(-\nabla^2_{(S)}\r)=-\int_0^\infty {dt\over t}\tr e^{t\nabla^2_{(S)}}\equiv -\int_0^\infty {dt\over t}K^{(S)}\l(t\r),
\end{equation}
where the trace of the Laplacian is taken over both the spin and the spacetime indices. Now for an arbitrary curved manifold $\mathcal{M}$, such expressions are prohibitively hard to compute, but symmetric spaces such as spheres and Euclidean AdS (i.e. hyperboloids) admit important simplifications due to an underlying group theoretic structure -- they can be realised as cosets of Lie groups (see for example \cite{Salam:1981xd}--\cite{Camporesi},\cite{David:2009xg}). The determinants of the Laplacian were explicitly evaluated for the symmetric, transverse-traceless (STT) fields in euclidean AdS in \cite{Gopakumar:2011qs} by the heat kernel method (see \cite{David:2009xg} for a more detailed exploration of AdS$_3$). We shall briefly recollect the main results, specialising to the case of $\ads$.

Firstly, euclidean $\ads$ is the symmetric space SO(5,1)/SO(5). The spin of a field over $\ads$ is given by the unitary irreducible representation (UIR) of the isotropy group SO(5) that is carried by the field. UIRs of SO(5) are labelled by the array
\begin{equation}
S=(s_1,s_2)\quad s.t. \quad s_1\geq s_2\geq 0.
\end{equation}
We shall also need UIRs of SO(5,1), which are labelled by the array
\begin{equation}
R=\l(i\lambda,m_1,m_2\r), \quad m_1\geq \vert m_2\vert.
\end{equation}
where $R$ contains $S$ if \cite{Ottoson,Camporesi}
\begin{equation}
\label{branching}
s_1\geq m_1\geq s_2\geq \vert m_2\vert.
\end{equation} 
The $m$'s and $s$'s must all simultaneously be integers or half-integers. Further, the $m$'s define an SO(4) UIR $\vec{m}=(m_1,m_2)$, and its conjugate representation $\check{\vec{m}}=(m_1,-m_2)$. With these ingredients, the heat kernel of a spin-$S$ particle on (a quotient of) $\ads$ is given by
\begin{equation}
\label{ads5intkernel}
K^{\l(S\r)}\l(\gamma,t\r)
=\frac{\beta}{2\pi}\sum_{k\in \bz}\sum_{\vec{m}}\int_0^\infty d\lambda\, \chi_{\lambda,\vec{m}} \l(\g^k\r) e^{tE_R^{(S)}},
\end{equation}
where $\chi_{\lambda,\vec{m}}$ is the Harish-Chandra character in the principal series of $SO\l(5,1\r)$, which has been evaluated \cite{HiraiChar} to be
\begin{equation}
\label{fullHCchar}
\chi_{\lambda,\vec{m}}\l(\beta,\phi_1,\phi_2\r)=
\frac{e^{-i\beta\lambda}\chi^{SO\l(4\r)}_{\vec{m}}\l(\phi_1,\phi_2\r)+e^{i\beta\lambda}\chi^{SO\l(4\r)}_{\check{\vec{m}}}\l(\phi_1,\phi_2\r)} {e^{-2\beta}\prod_{i=1}^2\vert e^\beta - e^{i\phi_i}\vert^2},
\end{equation}
the eigenvalue of the Laplacian $E_R^{(S)}$ is a function of $\lambda$
\begin{equation}
E_R^{(S)}= -\l(\lambda^2 +C_2(S)-C_2(\vec{m})+2^2\r),
\end{equation}
and $\g$ denotes the quotient of $\ads$ which corresponds to turning on a temperature $\beta$ along with angular momentum chemical potentials $\phi_1,\phi_2$ along the SO(4) Cartans. $C_2$ denotes the quadratic Casimir of the appropriate $SO$ group. The sum over $\vec{m}$ is the sum over all values of $m$ admitted by the branching rules \eqref{branching}. We refer the reader to \cite{Gopakumar:2011qs} for more details and explicit expressions.

There are many simplifications for the case of STT tensors. Firstly, as the tensor is completely symmetric, $S=(s,0)$, where $s$ is the rank of the tensor. Secondly, the branching rules \eqref{branching} simplify to give
\begin{equation}\label{STTbranching}
m_1=s,\,m_2=0.
\end{equation}
The partition function of a massless spin-$s$ particle was evaluated with these inputs in \cite{Gupta:2012np}. In particular, it was found that the expression for the partition function as a result of evaluating the path integral at one loop arranged itself in terms of transverse traceless tensors, and finally (see Eq. (2.29) of \cite{Gupta:2012np})
\begin{equation}
\label{STTpart}
\begin{split}
\log\ZZ^{(s)} = \sum_{m=1}^\infty {1\over m}&{e^{-m\beta \l(s+2\r)}\over \vert 1-e^{-m\l(\beta-i\phi_1\r)}\vert ^2 \vert 1-e^{-m\l(\beta-i\phi_2\r)}\vert ^2}\times \\ \times & \l[\chi_{s\over 2}\l(m\a_1\r)\chi_{s\over 2}\l(m\a_2\r)-\chi_{s-1\over 2}\l(m\a_1\r)\chi_{s-1\over 2}\l(m\a_2\r)e^{-m\beta}\r],\end{split}
\end{equation}
where we have expressed the $SO(4)$ character $\chi^{SO\l(4\r)}_{\l(s,0\r)}\l(m\phi_1,m\phi_2\r)$ as a product of $SU(2)$ characters $\chi_{s\over 2}\l(m\a_1\r)\chi_{s\over 2}\l(m\a_2\r)$, where $\alpha_1={\phi_1+\phi_2}$, and $\alpha_2={\phi_1-\phi_2}$. 

The reader will recognise \eqref{STTpart} as the expression for the multiparticle partition function in terms of the one-particle partition function $\YY$, where $\YY$ is the $SO(4,2)$ character evaluated over the short representation $[s+2,{s\over 2},{s\over 2}]$ \cite{Barabanschikov:2005ri,Dolan:2005wy}. Using the AdS/CFT correspondence \cite{Maldacena:1997re,Gubser:1998bc,Witten:1998qj}, if a CFT partition function contains a character of the representation $[s+2,{s\over 2},{s\over 2}]$, there must be bulk degrees of freedom giving rise to one-loop determinants over STT fields, as reviewed above. For example, given a long primary $[\D,{s\over 2},{s\over 2}]$ in the CFT, one can infer the presence of quadratic fluctuations in AdS giving rise to a one-loop determinant of the operator $-\nabla^2+m^2$, where $m^2=\l(\D-2\r)^2-s-4$. The corresponding heat kernel is given by
\begin{equation}
K^{[\Delta,{s\over 2},{s\over 2}]}\l(\g,t\r)=\frac{\beta}{\sqrt{\pi t}}\sum_{k=1}^\infty \frac{\chi_{s\over 2}\l(k\a_1\r)\chi_{s\over 2}\l(k\a_2\r)}{e^{-2k\beta}\vert e^{k\beta} - e^{ik{(\a_1+\a_2)\over 2}}\vert ^2 \vert e^{k\beta} - e^{ik{(\a_2-\a_1)\over 2}}\vert ^2}e^{-t\l(\D-2\r)^2}e^{-{k^2\beta^2\over 4t}}
\end{equation}
for the dual bulk fluctuations. It may be verified by doing the $t$ integral as in \cite{Gopakumar:2011qs} that this gives rise to the expected partition function. This forms the basis of the analysis of Section \ref{secansatz}.

\section{The Heat Kernel for Mixed Symmetry Fields}\label{secansatz}
In this section, we will compute the heat kernel for the $\ads$ degrees of freedom that correspond to primaries of mixed symmetry in the CFT. These correspond to representations $S$ of $SO(5)$ where $s_2\neq 0$, i.e.
\begin{equation}
S=(s_1,s_2)\quad s.t. \quad s_1\geq s_2 > 0.
\end{equation}
Representations of SO(5,1) $R$ that contain $S$ are determined by the branching rules \eqref{branching}. The main ingredient of this calculation will be the tensors for which some of the inequalities in the branching rules \eqref{branching} get saturated. In particular, that
\begin{equation}
\label{ttbranching}
m_1=s_1,\, \vert m_2\vert =s_2.
\end{equation}
We therefore have
\begin{equation}
R=\l(i\lambda,s_1,\pm s_2\r).
\end{equation}
The eigenvalues of the Laplacian, for such fields (which saturate the branching rules as above) transforming in $S=(s_1,s_2)$, are now given by
\begin{equation}\label{eigenvalues}
E_R^S=-\l(\lambda^2+s_1+s_2+4\r), \quad R=\l(i\lambda,s_1,\pm s_2\r).
\end{equation}
In what follows, we shall focus exclusively on such tensors in $AdS$. Using these expressions, we will now obtain a formula for the heat kernel for the Laplacian in AdS acting over such fields of mixed symmetry. This, by the above considerations, is given by
\begin{equation}
K^{(S)}\l(\gamma,t\r)={\beta\over 2\pi}\sum_{k \in \bz}\int_0^\infty d\lambda \l[ \chi_{\lambda,(s_1,s_2)}\l(\g^k\r)+\chi_{\lambda,(s_1,-s_2)}\l(\g^k\r)\r] e^{i E_R^S},
\end{equation}
where we have used the degeneracy of eigenvalues \eqref{eigenvalues} while carrying out the sum over $\hat{m}$ above. The sum 
\begin{equation}
\l[\chi_{\lambda,(s_1,s_2)}+\chi_{\lambda,(s_1,-s_2)}\r]\l(\g\r)= \frac{2\cos\beta\lambda\l[\chi^{SO(4)}_{(s_1,s_2)}\l(\phi_1,\phi_2\r) +\chi^{SO(4)}_{(s_1,-s_2)}\l(\phi_1,\phi_2\r)\r]}{e^{-2\beta}\vert e^\beta - e^{i\phi_1}\vert ^2 \vert e^\beta - e^{i\phi_2}\vert ^2}.
\end{equation}
Again, for later analysis it is more efficient to express the answer in terms of the $SU(2)\otimes SU(2)$ characters rather than the $SO(4)$ ones. The precise dictionary is
\begin{equation}
\chi^{SO(4)}_{(s_1,s_2)}\l(\phi_1,\phi_2\r)=\chi_{j_1}\l(\a_1\r)\chi_{j_2}\l(\a_2\r),
\end{equation}
where 
\begin{equation}
\label{dic}
\begin{split}
j_1={s_1+s_2 \over 2} &,\quad \alpha_1={\phi_1-\phi_2},\\
j_2={s_1-s_2 \over 2} &,\quad \alpha_2={\phi_1+\phi_2}.
\end{split}
\end{equation}
With these ingredients, the heat kernel corresponding to a TT tensor field of spin $S=(s_1,s_2)\equiv (j_1,j_2)$ is given by
\begin{equation}
K^{(j_1,j_2)}\l(\g,t\r)={\beta\over 2\pi}\sum_{k \in \bz}\int_0^\infty d\lambda \frac{2\cos k\beta\lambda\l(\chi_{j_1}\chi_{j_2} +\chi_{j_2}\chi_{j_1}\r)\l(k\a_1,k\a_2\r)}{e^{-2k\beta}\vert e^{k\beta} - e^{ik{(\a_1+\a_2)\over 2}}\vert ^2 \vert e^{k\beta} - e^{ik{(\a_2-\a_1)\over 2}}\vert ^2}e^{-t\l(\lambda^2+2j_1+4\r)},
\end{equation}
Note that we cannot naively apply this expression to symmetric tensor fields by setting $j_2=j_1$. This is because we have separately counted representations $(i\lambda,s_1,\pm s_2)$. If $s_2=0$, or equivalently $j_1=j_2$, these representations would be the same and we would count them only once. We will next apply these results to the case of massive and massless fields, but before doing so, let us perform the $\lambda$ integral to get an expression for the heat kernel purely in terms of the chemical potentials and the heat kernel proper time. We use the identity
\begin{equation}
\int_0^\infty 2 \cos k\beta\lambda e^{-t\lambda^2} d\lambda=\sqrt{{\pi\over t}}e^{-{k^2\beta^2\over 4t}}
\end{equation}
to write the heat kernel of $-\nabla^2$ as 
\begin{equation}\label{genkernel}
K^{(j_1,j_2)}\l(\g,t\r)={\beta\over \sqrt{\pi t}}\sum_{k=1}^\infty \frac{\l(\chi_{j_1}\chi_{j_2} +\chi_{j_2}\chi_{j_1}\r)\l(k\a_1,k\a_2\r)}{e^{-2k\beta}\vert e^{k\beta} - e^{ik{(\a_1+\a_2)\over 2}}\vert ^2 \vert e^{k\beta} - e^{ik{(\a_2-\a_1)\over 2}}\vert ^2}e^{-t\l(2j_1+4\r)}e^{-{k^2\beta^2\over 4t}},
\end{equation}
where 
\begin{equation}
\chi_{j_1}\chi_{j_2}(k\a_1 ,k\a_2)\equiv \chi_{j_1}(k\a_1)\chi_{j_2}(k\a_2).
\end{equation}
In the remainder of this section, we shall use \eqref{genkernel} as a building block for the bulk contributions that correspond to mixed symmetry primaries in the boundary CFT.
\subsection{The Heat Kernel for Massive Fields}
Consider now a long SO(4,2) representation of highest weight $[\D,j_1,j_2]\oplus [\D,j_1,j_2]$. This primary is dual to a massive field in the bulk. The  corresponding heat kernel is given by
\begin{equation}\label{longkernel}
K^{[\Delta,j_1,j_2]}\l(\g,t\r)=\frac{\beta}{\sqrt{\pi t}}\sum_{k=1}^\infty \frac{\l(\chi_{j_1}\chi_{j_2} +\chi_{j_2}\chi_{j_1}\r)\l(k\a_1,k\a_2\r)}{e^{-2k\beta}\vert e^{k\beta} - e^{ik{(\a_1+\a_2)\over 2}}\vert ^2 \vert e^{k\beta} - e^{ik{(\a_2-\a_1)\over 2}}\vert ^2}e^{-t\l(\D-2\r)^2}e^{-{k^2\beta^2\over 4t}}.
\end{equation}
This corresponds to the heat kernel of the operator $-\nabla^2+m^2$ evaluated on the tensor fields \eqref{ttbranching}, where $m^2+2j_1+4=\l(\D-2\r)^2$. Then
\begin{equation}
\log\ZZ=-{1\over 2}\log\det\l(-\nabla^2+m^2\r)={1\over 2}\int_0^\infty {dt\over t}K^{[\Delta,j_1,j_2]}\l(\g,t\r).
\end{equation}
We carry out the $t$ integral using equation 6.1 of \cite{Gopakumar:2011qs}
\begin{equation}\label{timeintegral}
{1\over 2\sqrt{\pi}}\int_0^\infty {dt\over t^{3\over 2}} e^{-{\a^2\over 4t}-\beta^2t}={1\over\a}e^{-\a\beta}.
\end{equation}
We finally find that
\begin{equation}
\log\ZZ=\sum_{k=1}^\infty \frac{\l(\chi_{j_1}\chi_{j_2} +\chi_{j_2}\chi_{j_1}\r)\l(k\a_1,k\a_2\r)}{\vert 1 - e^{-k\beta}e^{ik{(\a_1+\a_2)\over 2}}\vert ^2 \vert 1 - e^{-k\beta}e^{ik{(\a_2-\a_1)\over 2}}\vert ^2} e^{-k\beta\D}.
\end{equation}
To summarise, we have 
\begin{equation}\label{convert1}
\log\mathcal{Z}_{[\D,j_1,j_2]\oplus[\D,j_2,j_1]}={1\over 2}\int_0^\infty {dt\over t} K^{[\D,j_1,j_2]}.
\end{equation}
\subsection{The Heat Kernel for Mixed Symmetry Massless Fields}
Massless fields belong to short representations of the conformal group. The partition function that corresponds to these fields is the character of a short representation with highest weight $[j_1+j_2+2,j_1,j_2]$, i.e.
\begin{equation}
\log\mathcal{Z}_{[j_1+j_2+2,j_1,j_2]}=\sum_{k=1}^\infty {1\over k}\l(\chi_{[j_1+j_2+2,j_1,j_2]}-\chi_{[j_1+j_2+3,j_1-{1\over 2},j_2-{1\over 2}]}\r)\l(k\beta ,k\alpha_1 ,k\alpha_2\r),
\end{equation}
where the characters on the right-hand side are the characters over long representations of SO(4,2). We can therefore write
\begin{equation}
\log\mathcal{Z}_{[j_1+j_2+2,j_1,j_2]}={1\over 2}\int_0^\infty {dt\over t} \l(K^{[j_1+j_2+2,j_1,j_2]}-K^{[j_1+j_2+3,j_1-{1\over 2},j_2-{1\over 2}]}\r)\l(\beta ,\alpha_1 ,\alpha_2\r),
\end{equation}
in the notation of \eqref{longkernel}. 
\section{From the CFT Partition Function to the AdS Heat Kernel}\label{secanswers}
We now organise the full CFT partition function into the form of a heat kernel in $\ads$ for a theory which has only long representations in its spectrum. Remarkably, however it turns out that the final answer \eqref{longfinal} thus obtained is unchanged when short multiplets \cite{Mack:1975je,Minwalla:1997ka} are included. This is demonstrated in Appendix \ref{appshortkernel}.

Suppose the CFT has operators with quantum numbers $[\D,j_1,j_2]$ appearing $N_{[\D,j_1,j_2]}$ times. The one-particle partition function is a sum of SO(4,2) characters evaluated over the modules generated by these primaries.
\begin{equation}\label{partfny}
\mathcal{Y}(q,a,b)= \sum_{\D,j_1,j_2} N_{[\D,j_1,j_2]}{q^{\D}\chi_{j_1}(a)\chi_{j_2}(b)\over \prod_{i=1}^4 (1-qx_i)},
\end{equation}
The (multi-particle) partition function of the theory is then obtained by exponentiating the one-particle partition function.
\begin{equation}
\log\mathcal{Z}=\sum_{k=1}^\infty {1\over k}\sum_{\D,j_1,j_2} N_{[\D,j_1,j_2]}\chi_{[\D,j_1,j_2]}\l(q^k,a^k,b^k\r).
\end{equation}
where we have introduced notation
\begin{equation}
q=e^{-\beta},\,a = e^{i\a_1} ,\,b=e^{i\a_2}.
\end{equation}
Shortly we will also define
\begin{equation}
x_1 =\sqrt{ab},x_2=\sqrt{a\bar{b}},x_3=\sqrt{\bar{a}b},x_4=\sqrt{\bar{a}\bar{b}}. 
\end{equation}
In what follows, it is useful to treat symmetric and mixed-symmetric tensors on a different footing, i.e. sum up the $j_1=j_2$ and $j_1\neq j_2$ contributions separately. We then have 
\begin{equation}
\log\mathcal{Z}=\sum_{k=1}^\infty {1\over k}\sum_{\D,j} N_{[\D,j,j]}\chi_{[\D,j,j]}+\sum_{k=1}^\infty {1\over k}\cdot{1\over 2}\sideset{}{'}\sum_{\D,j_1,j_2} N_{[\D,j_1,j_2]}\l(\chi_{[\D,j_1,j_2]}+\chi_{[\D,j_2,j_1]}\r).
\end{equation}
The prime over the second sum reminds us that in this sum, $j_1\neq j_2$. The factor of half in the second term is from the fact that this sum counts each $(j_1,j_2)$ pair twice. The dependence on $(q^k,a^k,b^k)$ is implicit. We have imposed the condition that $N_{[\D,j_1,j_2]}=N_{[\D,j_2,j_1]}$ to club terms together in the second sum.

We will now reinterpret, as per \eqref{longkernel}, each Verma module character above as arising from a heat kernel in $\ads$. Using \eqref{convert1}, we have
\begin{equation}\label{kernelsum}
\log\mathcal{Z}={1\over 2}\int_0^\infty {dt\over t}\l( \sum_{\D,j} N_{[\D,j,j]}K^{[\D,j,j]}+{1\over 2}\sideset{}{'}\sum_{\D,j_1,j_2} N_{[\D,j_1,j_2]}K^{[\D,j_1,j_2]}\r).
\end{equation}
We will now evaluate the sums over $\D,j_1,j_2$. To do so, the following identity is useful
\begin{equation}\label{intidentity}
e^{-t\l(\D-2\r)^2}=\sqrt{{1\over 4\pi t}}\int_{-\infty}^{\infty}dy e^{-{y^2\over 4 t}+iy\l(\D-2\r)}.
\end{equation}
This follows from evaluating the Gaussian integral on the right-hand side. The heat kernel formulae in our new notations are
\begin{equation}\label{newkernel}
\begin{split}
K^{[\Delta,j,j]}\l(\g,t\r)&=\sum_{k=1}^\infty {\beta\over 2\pi t}\int_{-\infty}^\infty dy e^{-{y^2+k^2\beta^2\over 4t}} e^{-2iy} \frac{q^{2k} \chi_{j}(a^k)\chi_{j}(b^k) }{\prod_{i=1}^4\l(1-q^kx^k_i\r)}e^{i\D y}\\
K^{[\Delta,j_1,j_2]}\l(\g,t\r)&=\sum_{k=1}^\infty {\beta\over 2\pi t}\int_{-\infty}^\infty dy e^{-{y^2+k^2\beta^2\over 4t}} e^{-2iy} \frac{q^{2k}\l(\chi_{j_1}(a^k)\chi_{j_2}(b^k) +\chi_{j_2}(a^k)\chi_{j_1}(b^k)\r)}{\prod_{i=1}^4\l(1-q^kx^k_i\r)}e^{i\D y}.
\end{split}
\end{equation}
We will now use these expressions to evaluate \eqref{kernelsum}. As is apparent, most of \eqref{newkernel} does not depend on $\D,j_1,j_2$ and factors out of the sum. The sum that we essentially have to evaluate is
\begin{equation}\begin{split}
\sum_{\D,j} N_{[\D,j,j]}e^{i\D y}\chi_{j}(a^k)\chi_{j}(b^k)+{1\over 2}\sideset{}{'}\sum_{\D,j_1,j_2} &N_{[\D,j_1,j_2]}e^{i\D y} \l(\chi_{j_1}(a^k)\chi_{j_2}(b^k) +\chi_{j_2}(a^k)\chi_{j_1}(b^k)\r)\\&=
\sum_{\D,j_1,j_2} N_{[\D,j_1,j_2]}e^{i\D y}\chi_{j_1}(a^k)\chi_{j_2}(b^k).\end{split}
\end{equation}
We can now use the definition \eqref{partfny} (replacing $q$ by $e^{iy}$) to write this as
\begin{equation}
\mathcal{Y}\l(e^{i y},a^k,b^k\r)\prod_{i=1}^4\l(1-e^{i y}x^k_i\r).
\end{equation}
We therefore find that
\begin{equation}
\log\mathcal{Z}\simeq \sum_{k=1}^\infty {\beta\over 2\pi t}\int_{-\infty}^\infty dy e^{-{y^2+k^2\beta^2\over 4t}} e^{-2iy}q^{2k} \mathcal{Y}\l(e^{iy},a^k,b^k\r)\prod_{i=1}^4{\l(1-e^{i y}x^k_i\r)\over \l(1-q^kx^k_i\r)}.
\end{equation}
Including the integral over $t$ explicitly, we find that
\begin{equation}\label{longfinal}
\log\mathcal{Z}= \sum_{k=1}^\infty\int_0^\infty dt{\beta\over 4\pi t^2}\int_{-\infty}^\infty dy e^{-{y^2+k^2\beta^2\over 4t}} e^{-2iy}q^{2k} \mathcal{Y}\l(e^{iy},a^k,b^k\r)\prod_{i=1}^4{\l(1-e^{iy}x^k_i\r)\over \l(1-q^kx^k_i\r)}.
\end{equation}
This is an expression for the multi-particle partition function of the $\ads$ theory in terms of the heat kernel time $t$ and the single-particle partition function of its dual CFT. We remind the reader that the single-particle partition function $\YY$ has been very explicitly computed for a free, planar CFT by enumerating the gauge invariant operators in the CFT spectrum \cite{Sundborg:1999ue,Aharony:2003sx}.\footnote{Taking the free limit in the CFT side corresponds to the limit in which the string theory on AdS becomes tensionless. This is of course a very non-trivial limit of string theory about which much remains to be understood. However, from the AdS/CFT correspondence, the spectrum of string theory should still match with the spectrum of conformal primaries of the CFT, and our analysis would still be valid. At generic values of the coupling, the spectrum of string theory is also implicitly known through TBA, see \cite{Beisert:2010jr} for a review. We thank Arkady Tseytlin for discussions regarding these points.}
\section{Conclusions}
In this note, we reinterpreted the partition function of a free CFT on S$^3\otimes$S$^1$ in terms of the heat kernel corresponding to a free string in AdS$_5$ with all chemical potentials turned on. The main ingredients of this calculation were the heat kernel results of \cite{Gopakumar:2011qs}. The answer was expressed as a convolution of the one-particle partition function of the dual CFT. This would correspond to the one-loop string path integral in $\ads$ \textit{after} the level matching condition is imposed. In the future, we would look to examine how this answer might be extended to recover the full string path integral before level matching is imposed. This is work in progress, and we hope to report on it soon.

Finally, we mention that the results of \cite{Gopakumar:2011qs} extend to arbitrary dimensional hyperboloids, though the odd-dimensional case is perhaps the nicest. Corresponding expressions for the characters of the conformal group are also available \cite{Dolan:2005wy}. Therefore, the analysis presented here extends straightforwardly to AdS/CFT dualities in other dimensions as well.
\section*{Acknowledgements} 
We would like to thank Archisman Ghosh and especially Rajesh Gopakumar for several very helpful discussions and for encouragement to publish these results. We would also like to thank Arkady Tseytlin and Shahin Mammadov for correspondence. Additionally, we thank the Harish-Chandra Research Institute for support in the form of a Senior Research Fellowship while this work was initiated, and The Institute of Mathematical Sciences, Chennai for hospitality in the course of this work. More generally, we express gratitude to the people of India for their generous support to research in theoretical sciences.
\section*{Appendix}
\appendix
\section{Incorporating Short Multiplets}\label{appshortkernel}
In this section we will examine how the formula \eqref{longfinal} extends to the case where the spectrum of the theory has short multiplets of SO(4,2). To do so, we will consider a toy example where the theory has a short multiplet $[2m+2,m,m]$ that appears  $N_{\lbrace m\rbrace}$ times, another short multiplet $[\ell_1+\ell_2+2,\ell_1,\ell_2]\oplus [\ell_1+\ell_2+2,\ell_2,\ell_1]$ that appears $N_{\lbrace \ell\rbrace}$ times. All other multiplets are long. Given this spectrum
\begin{equation}\begin{split}
\log\mathcal{Z}= \sum_{k=1}^\infty {1\over k}&\l(\sideset{}{'}\sum_{\D,j} N_{[\D,j,j]}\chi_{[\D,j,j]}+N_{\lbrace m\rbrace}\l(\chi_{[2m+2,m,m]}-\chi_{[2m+3,m-{1\over 2},m-{1\over 2}]}\r)\r)\\+\sum_{k=1}^\infty {1\over k}&\l(\sideset{}{''}\sum_{\D,j_1,j_2} N_{[\D,j_1,j_2]}\chi_{[\D,j_1,j_2]}+N_{\lbrace \ell\rbrace}\l(\chi_{[\ell_1+\ell_2+2,\ell_1,\ell_2]}\r.\r.-\chi_{[\ell_1+\ell_2+3,\ell_1-{1\over 2},\ell_2-{1\over 2}]}\\  &\l.\l.+\chi_{[\ell_1+\ell_2+2,\ell_2,\ell_1]}-\chi_{[\ell_1+\ell_2+3,\ell_2-{1\over 2},\ell_1-{1\over 2}]}\r)\vphantom{\sideset{}{''}\sum_{\D,j_1,j_2}}\r).\end{split}
\end{equation}
The prime in the first term is to indicate that the short representation $[2m+2,m,m]$ is not summed over. The double-prime on the second sum is to indicate that the short representation involving the $\ell$s as well as the terms where $j_1=j_2$ are not summed in this. This then simplifies to 
\begin{equation}\begin{split}
\log\mathcal{Z}=&\sum_{k=1}^\infty {1\over k}\sum_{\D,j} N_{[\D,j,j]}\chi_{[\D,j,j]}+\sum_{k=1}^\infty {1\over k}\cdot{1\over 2}\sideset{}{'}\sum_{\D,j_1,j_2} N_{[\D,j_1,j_2]}\l(\chi_{[\D,j_1,j_2]}+\chi_{[\D,j_2,j_1]}\r)\\&- \sum_{k=1}^\infty {1\over k}\l(N_{\lbrace m\rbrace}\chi_{[2m+3,m-{1\over 2},m-{1\over 2}]}+N_{\lbrace \ell\rbrace}\l(\chi_{[\ell_1+\ell_2+3,\ell_1-{1\over 2},\ell_2-{1\over 2}]}+\chi_{[\ell_1+\ell_2+3,\ell_2-{1\over 2},\ell_1-{1\over 2}]}\r)\r).\end{split}
\end{equation}
Using these results we can write
\begin{equation}\begin{split}
\log\mathcal{Z}\simeq \sum_{\D,j} &N_{[\D,j,j]}K^{[\D,j,j]}+{1\over 2}\sideset{}{'}\sum_{\D,j_1,j_2} N_{[\D,j_1,j_2]}K^{[\D,j_1,j_2]}-N_{\lbrace m\rbrace}K^{[2m+2,m,m]}\\&-N_{\lbrace \ell\rbrace}K^{[\ell_1+\ell_2+3,\ell_1-{1\over 2},\ell_2-{1\over 2}]},\end{split}
\end{equation}
where we have adopted the notation 
\begin{equation}\label{convert2}
\log\mathcal{Z}\simeq K,
\end{equation}
where which we mean that K is the heat kernel which integrates to the partition function $\mathcal{Z}$ via \eqref{timeintegral}. Now the rest of the calculation proceeds exactly as before. In particular, we use \eqref{intidentity} and \eqref{newkernel} to show that the sum we actually need to do is
\begin{equation}\begin{split}
&\sum_{\D,j} N_{[\D,j,j]}e^{i\D y}\chi_{j}(a^k)\chi_{j}(b^k)+{1\over 2}\sideset{}{'}\sum_{\D,j_1,j_2} N_{[\D,j_1,j_2]}e^{i\D y} \l(\chi_{j_1}(a^k)\chi_{j_2}(b^k) +\chi_{j_2}(a^k)\chi_{j_1}(b^k)\r)\\&-N_{\lbrace m\rbrace}e^{i(2m+3)y}\chi_{m-{1\over 2}}(a^k)\chi_{m-{1\over 2}}(b^k)-N_{\lbrace\ell\rbrace}e^{i(\ell_1+\ell_2+3)y}\l(\chi_{\ell_1-{1\over 2}}(a^k)\chi_{\ell_2-{1\over 2}}(b^k)+\chi_{\ell_1-{1\over 2}}(a^k)\chi_{\ell_2-{1\over 2}}(b^k)\r)\\&\equiv\sum_{\D,j_1,j_2} N_{[\D,j_1,j_2]}e^{i\D y} \l(\chi_{j_1}(a^k)\chi_{j_2}(b^k) +\chi_{j_2}(a^k)\chi_{j_1}(b^k)\r)-N_{\lbrace m\rbrace}e^{i(2m+3)y}\chi_{m-{1\over 2}}(a^k)\chi_{m-{1\over 2}}(b^k)\\&-N_{\lbrace\ell\rbrace}e^{i(\ell_1+\ell_2+3)y}\l(\chi_{\ell_1-{1\over 2}}(a^k)\chi_{\ell_2-{1\over 2}}(b^k)+\chi_{\ell_1-{1\over 2}}(a^k)\chi_{\ell_2-{1\over 2}}(b^k)\r).
\end{split}\end{equation}
This simplifies on using the expression for the one-particle partition function to find that this is still given by
\begin{equation}
\mathcal{Y}\l(e^{i y},a^k,b^k\r)\prod_{i=1}^4\l(1-e^{i y}x^k_i\r).
\end{equation}
This implies that the final result for $\log\mathcal{Z}$ is still given by \eqref{longfinal}.
\section{A Consistency Check}
Equation \eqref{longfinal} expresses the second quantised partition function of the $\ads$ string in terms of the heat kernel time $t$ and the single-particle partition function of the dual CFT. As a check on this expression, we will exchange the integrals to perform the $t$ integral first, and then carry out the $y$ integral. The following identity is useful.
\begin{equation}
\int_0^\infty {dt\over t^2}e^{-{(y^2+k^2\beta^2)\over 4t}}={4\over y^2+k^2\beta^2}.
\end{equation}
After carrying out the $t$ integral as above, we find that
\begin{equation}
\log\mathcal{Z}= \sum_{k=1}^\infty {\beta\over 4\pi}\int_{-\infty}^\infty dy {4\over y^2+k^2\beta^2} e^{-2iy}q^{2k} \mathcal{Y}\l(e^{iy},a^k,b^k\r)\prod_{i=1}^4{\l(1-e^{iy}x^k_i\r)\over \l(1-q^kx^k_i\r)}.
\end{equation}
The integral over $y$ is may be carried out by closing the contour of integration over the upper half-plane in $y$ to pick up the pole at $y=+ik\beta$. We therefore find that
\begin{equation}
\log\mathcal{Z}= \sum_{k=1}^\infty {\beta\over\pi} 2\pi i {1\over 2ik\beta} e^{-2i(ik\beta)}q^{2k} \mathcal{Y}\l(e^{i(ik\beta)},a^k,b^k\r)\prod_{i=1}^4{\l(1-e^{i(ik\beta)}x^k_i\r)\over \l(1-q^kx^k_i\r)}.
\end{equation}
Recollecting that $q=e^{-\beta}$, we find that
\begin{equation}
\log\mathcal{Z}= \sum_{k=1}^\infty {1\over k} \mathcal{Y}\l(q^k,a^k,b^k\r),
\end{equation}
which is the familiar expression for the second quantised partition function as the exponential of the one-particle partition function.

\end{document}